\documentclass[prl,twocolumn,aps,showpacs]{revtex4}
\usepackage{graphicx}
\usepackage{bm}
\usepackage{amssymb} 
\begin{document}

\title{Scattering of plasmons at the intersection
of two nanotubes: Implications for tunnelling}

\author{V. V. Mkhitaryan, Y. Fang, J. M. Gerton, E. G. Mishchenko,
and M. E. Raikh}

\affiliation{ Department of Physics, University of Utah, Salt Lake
City, UT 84112}

\begin{abstract}
We study theoretically the plasmon scattering at the intersection
of two metallic carbon nanotubes. We demonstrate that for a small
angle of crossing, $\theta \ll 1$, the transmission coefficient is
an {\em oscillatory} function of $\lambda/\theta$, where $\lambda$
is the interaction parameter of the Luttinger liquid in an
individual nanotube. We calculate the tunnel density of states,
$\nu(\omega,x)$, as a function of energy, $\omega$, and  distance,
$x$, from the intersection. In contrast to a single nanotube, we
find that, in the geometry of crossed nanotubes, conventional
``rapid'' oscillations in $\nu(\omega,x)$ due to the plasmon
scattering acquire an aperiodic ``slow-breathing'' envelope which
has $\lambda/\theta$ nodes.

\end{abstract}
 \maketitle

\noindent{\em Introduction.} By now, observation of
Luttinger liquid in 1D systems has been reported for single-walled
carbon nanotubes \cite{bockrath,yao99,postma,ishii03,gao,dai} and
GaAs-based semiconductor
wires \cite{Yacoby08}.
Conclusions about Luttinger liquid behavior have been drawn from
analysis of the data, which can be divided into two groups: ({\em
i}) power-law, $\propto (\mbox {max}\{V,T\})^{\alpha}$, behavior
of tunnel or source-drain conductance
\cite{bockrath,yao99,postma,ishii03,gao,dai}, where parameter
$\alpha$ is the measure of deviation from the Fermi liquid
behavior, and ({\em ii}) momentum-resolved tunnelling in a
parallel magnetic field \cite{Yacoby08}.

On the conceptual level, the difference between the techniques
({\em i}) and ({\em ii}) is that ({\em i}) probes  a {\em
single-point} Green function, $\mathcal{ G}(x,x,\omega)$, while
({\em ii}), by mapping $\int dx \int dx^{\prime}
\mathcal{G}(x,x^{\prime},\omega)\exp[-iq_{\scriptscriptstyle
B}(x-x^{\prime})]$, with $q_{\scriptscriptstyle B}$ proportional
to applied field, yields information about {\em two-point} Green
function, and thus is more informative.

\begin{figure}[t]
\centerline{\includegraphics[width=60mm,angle=0,clip]{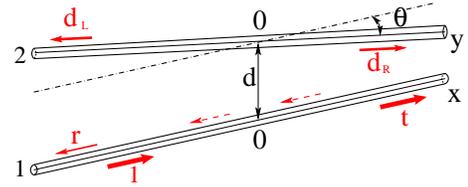}}
\caption{(Color online)
Intersecting nanotubes, separated by a distance, $d$; the angle of
intersection is $\theta$. Directions of incident, reflected, $r$,
transmitted, $t$, and deflected, $d_L$, $d_R$, plasmon waves are
shown with solid red arrows. Dashed arrows illustrate two
contributions to the reflected wave.}
\end{figure}
With regard to quantitative determination of the Luttinger liquid
parameter, $g$, which is related to $\alpha$ as \cite{kane97}
$\alpha=(g^{-1}+g-2)/8$, it is desirable to identify an effect,
which would depend on $g$ stronger than a power law. An example of
such an effect was given by Ussishkin and Glazman in Ref.
\onlinecite{glazman04}, where, due to electron backscattering, $g$
appears in the {\em argument of sine}; this sine describes the
amplitude modulation of the probe-induced Friedel oscillations
\cite{gogolin95}, $\propto \cos(2k_{\scriptscriptstyle F}x)$ in
the local density of states; $k_{\scriptscriptstyle F}$ is the
Fermi momentum.

In the present paper we demonstrate that the geometry of the
crossed  nanotubes (see Fig.~1) offers a qualitatively new
manifestation of the Luttinger liquid behavior. In particular, the
oscillatory dependence on $g$, similar to that in  Ref.
\onlinecite{glazman04}, emerges in the geometry of crossed
nanotubes even without {\em electron} backscattering
\cite{egger98,fuhrer00,gao,crossedjunc}. More precisely, we show
that, in this geometry, the envelope, ``breathing'' with $g$,
modulates not $\cos(2k_{\scriptscriptstyle F}x)$ oscillations, but
much slower oscillations resulting from the {\em plasmon
backscattering}.

There is an important difference between scattering of plasmons
and electrons: for an obstacle bigger than $k_{\scriptscriptstyle
F}^{-1}$ electron scattering is exponentially suppressed, while
plasmon scattering is efficient as long as the size of the
obstacle does not exceed the {\em plasmon} wavelength. This
scattering gives rise to the oscillations of the local density of
states $\delta \nu(\omega,x) \propto \cos(2\omega
x/v_{\scriptscriptstyle F})$ where $v_{\scriptscriptstyle F}$ is
the Fermi velocity. It is these oscillations that acquire a
breathing envelop in the geometry of crossed nanotubes, Fig. 2.
Our main finding is that,  with regard to this modulation, making
the crossing angle $\theta$ small, effectively {\em enhances} the
Luttinger liquid parameter. To describe this enhancement
quantitatively, we first consider an auxiliary problem of plasmon
scattering at the intersection and later utilize it for the
calculation of $\delta\nu(\omega,x)$.

\begin{figure}[t]
\centerline{\includegraphics[width=80mm,angle=0,clip]{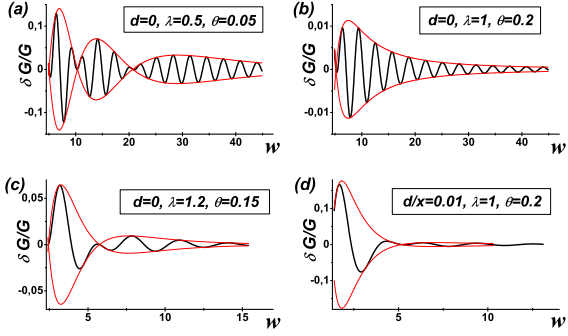}}
\vspace{-0.5cm} \caption{(Color online) Oscillating corrections to
the tunnelling conductance is plotted from Eq.~(\ref{mr}) versus
dimensionless bias $w =Vx/s$, where $x$ is the distance from the
intersection.
Periodic oscillations (black) are modulated by "breathing"
envelope (red), with "period" determined by $\lambda/\theta$,
where $\lambda$ is the interaction parameter; (d) illustrates
suppression of oscillations at finite separation, $d=0.01x$,
between the nanotubes.}
\end{figure}

\noindent{\em Plasmon scattering at the intersection.} Assume that
$d$ is the minimal distance between the nanotubes. Even in the
absence of electron tunnelling, a plasmon propagating towards
$x=0$ in the nanotube 1 can: ({\em i}) pass $x=0$ (transmission);
({\em ii}) excite a plasmon in the nanotube 2, which propagates
away from the intersection  $x=0$ either to the left or to the
right (deflection); ({\em iii}) get reflected. Incorporating the
plasmon scattering into the formalism of the Luttinger liquid
gives rise to the breathing envelope in Fig. 2.  The underlying
reason is that the interaction between the tubes strengthens
towards intersection \cite{maslov}. This leads to the $x$-{\em
dependent splitting} of velocities in each tube. The resulting
$x$-dependent phase accumulation near the intersection translates
into nontrivial dependence of $\delta\nu(\omega,x)$. Moreover, the
phase accumulation increases rapidly with decreasing angle
$\theta$, thus simulating the enhancement of the Luttinger
parameter.

\noindent{\it Collective modes of intersecting nanotubes.} As a
result of long-range interaction, $e^2\int \frac{dx dx^{\prime}}
{|x-x^{\prime}|} \frac{\partial u(x)}{\partial x}\frac{\partial
u(x^\prime)} {\partial x^\prime}$, where $u(x)$ is the
displacement of the electron position from the equilibrium, the
plasmon spectrum of an individual tube is $\omega(q)=qs(q)$ with
velocity \cite{kane97} $s=v_{\scriptscriptstyle F}(1+\lambda\ln
(qr))^{1/2}$. Here $r$ is the nanotube radius and
$\lambda=8e^2/(\pi\hbar v_{\scriptscriptstyle F})$ is the
interaction constant \cite{kane97}. Following Ref.
\onlinecite{kane97}, we neglect the relative change of $\ln(qr)$.
At a given frequency, $\omega$, displacement $u(x)$ is the
eigenmode, $\hat{D}\{u(x)\}=\left(\omega^2/v_{\scriptscriptstyle
F}^2\right)u(x)$, of the operator
\begin{eqnarray}
\label{opD} \hat{D}\{f\}=-\frac{\partial^2 }{\partial x^2}f(x)-
\lambda\frac{\partial^2 }{\partial x^2}\! \int_{-\infty}^\infty
\frac{dy}{|x-y|}f(y).
\end{eqnarray}
For two crossed nanotubes, the eigenmodes are described by the
system of two coupled equations,
\begin{eqnarray}
\label{ce1} &&\left(\omega^2/v_{\scriptscriptstyle F}^2\right)u_{1,2}(x)=
\hat{D}\{u_{1,2}(x)\} +\hat{F}\{u_{2,1}(x)\},\\
&&\hspace{-1cm}\hat{F}\{f\}=-\lambda\frac{\partial
}{\partial y}\!\!\int_{-\infty}^\infty \!\!
\frac{dx}{\sqrt{d^2+x^2+y^2-2xy\cos\theta}}\frac{\partial
f(x)}{\partial x}.\label{opF}
\end{eqnarray}
The operator $\hat{F}$ has a meaning of longitudinal force created
by the density
fluctuation, $\partial u_1(x)/\partial x$, in the nanotube $1$, at
point $y$ of the nanotube $2$. The scattering problem corresponds to
the solution of Eqs.~(\ref{ce1})
which has the following asymptotes at large $x$ and $y$:
\begin{eqnarray}
\label{ScatProblem}
&&u_1(x)\big |_{x\rightarrow-\infty}=e^{ikx}+re^{-ikx},\quad
u_1(x)\big |_{x\rightarrow\infty}=te^{ikx},\nonumber\\
&&u_2(y)\big |_{y\rightarrow-\infty}=d_Le^{-iky},\quad u_2(y)\big
|_{y\rightarrow\infty}=d_Re^{iky}.
\end{eqnarray}
\noindent{\it  Born approximation.} For small $\lambda$, the
elements of scattering matrix can be found in the Born
approximation in momentum space. To the first order in $\lambda$,
only $d_R$ and $d_L$ are non-zero. They are given by matrix
elements of the operator $\hat{F}$, Eq.~(\ref{opF}), namely,
$d_R(k)=(i/2k)\hat{F}_{k,k}$ and $d_L(k)=-(i/2k)\hat{F}_{-k,k}$.
Analytical expression for $\hat{F}_{p,q}$ is
\begin{eqnarray}\label{me}
&&\hspace{-0.3cm}\hat{F}_{p,q}=2\pi \lambda\,\frac{\,pq\,
 e^{-\frac{d}{\sin\theta}
\left(p^2+q^2-2pq\cos\theta\right)^{1/2}}}
{\left(p^2+q^2-2pq\cos\theta\right)^{1/2}}.
\end{eqnarray}
This leads to the final result for deflection coefficients
\begin{eqnarray}\label{fd}
d_R= i\pi \lambda\frac {\,\,e^{-\frac{kd}{\cos
(\theta/2)}}}{2\sin\left(\frac\theta 2\right)},\quad d_L= -i\pi
\lambda \frac{\,\,e^{-\frac{kd}{\sin
(\theta/2)}}}{2\cos\left(\frac\theta 2\right)}.
\end{eqnarray}
An apparent consequence of Eq.~(\ref{fd}) is that deflection is
exponentially small when the plasmon wavelength is $\ll d$. Less
obvious is that, for $kd<1$ and small $\theta$, coefficients $d_R$
and $d_L$ can differ {\it exponentially}. This is because the
exponent, $\exp(-2kd/\theta)$, in $d_L$ can be small if $kd$ is
small. Noteworthy, in the long-wavelength limit, $kd\ll\theta$, we
still have $d_R/d_L\approx 1/\theta\gg1$. The underlying reason is
that $ d_L$ corresponds to the wave which travels almost in the
opposite direction to the incident wave, while $d_R$ travels
almost along the incident wave. From Eq.~(\ref{fd}) we conclude
that the Born approximation applies at $\lambda\ll\theta$.

The reflection coefficient, $r(k)$, in the second Born
approximation, is expressed via the matrix elements Eq.~(\ref{me})
\begin{eqnarray}\label{rc}
r(k)=\frac1 {4i\pi k}\int_{-\infty}^{\infty}\,
\frac{dp}{p^2-k^2-i\epsilon}\hat{F}_{-k,p}\hat{F}_{p,k}.
\end{eqnarray}
The integral in Eq.~(\ref{rc}) is the sum, $(2\pi
\lambda)^2k(I_1+iI_2)$, of contributions from the poles $p=\pm k$
and the principal value, which can be cast in the form
\begin{eqnarray}\label{2I}
&&\hspace{-0.4cm}I_1=P\int\frac{dp\, p^2}{p^2-k^2}
\frac{e^{-\frac{d}{\sin\theta} (p^2+k^2-2pk\cos\theta)^{1/2}
}}{\bigl[(p^2+k^2)^2-4p^2k^2\cos^2\theta\bigr]^{1/2}}\nonumber\\
&&\times e^{-\frac{d}{\sin\theta}
(p^2+k^2+2pk\cos\theta)^{1/2}},\nonumber\\
&&\hspace{-0.4cm}I_2=\frac{\pi }{2\sin\theta}
e^{-kd\left[\frac1{\cos(\theta/2)}+\frac1{\sin(\theta/2)}\right]}.
\end{eqnarray} For short wavelengths, $kd\gg1$, the
dependence $r(k)$ is dominated by the integral $I_1\approx
-(\lambda^2/2)(kd\sin\theta/\pi)^{-3/2}e^{-2kd/\sin\theta}$. In
the long-wavelength limit, $kd\ll1$, one can replace the exponent
$e^{-kd/\sin\theta}$ by $1$. In what follows, we will focus on
small $\theta$, where $d_L$ and $r$ diverge. Note that the pole
contribution in Eq.~(\ref{2I}) diverges for $\theta\rightarrow 0$
much stronger that the principal value contribution, which is
$\propto \ln(1/\theta)$, so that $r\approx
\pi^2\lambda^2/2\theta$. We also notice that in the small-$\theta$
domain, the relation $r\approx d_Ld_R$ holds. This relation can be
understood from the following reasoning.

There are two contributions to the reflected wave in the second
Born approximation. ({\it i}) The wave deflected into the second
tube to the right with the amplitude (solid arrow in Fig.~1),
undergoes a secondary deflection back into the first tube (dashed
arrow in Fig.~1) with amplitude $d_L$. ({\it ii}) The wave
deflected into the second tube to the left, $d_L$, is subsequently
deflected back into the first tube with the amplitude $d_R$,
Fig.~1. The sum of the two contributions amounts to
$r=(c_1+c_2)d_Ld_R$. Remarkably, both numerical factors $c_1$ and
$c_2$ are equal to $1/2$. This is a consequence of a strong
difference in distances at which formation of the primary left-
and right-deflected waves takes place. The wave $d_L$ is formed
within $\sim 1/k$ from the intersection, while the wave, $d_R$, is
formed within a much broader interval $\sim 1/(k\theta)$.
Therefore, in second tube, at some distance $y$ from the
intersection, such that $1/k \ll y \ll 1/(k\theta)$, the amplitude
of the left-deflected wave is already $d_L$, while  the amplitude
of the right-deflected wave is only $\frac12d_R$. Subsequent
formation of the contribution ({\it ii}) occurs at $y\sim
1/(k\theta)$, so that the corresponding amplitude is
$\left(\frac12d_R\right)d_L$, i.e., $c_2=1/2$. On the other hand,
formation of the contribution ({\it i}) takes place only over {\it
negative} $-1/(k\theta)<y<0$, and thus results in
$d_L\left(\frac12d_R\right)$, i.e., $c_1=1/2$.

\noindent{\it Semiclassical description}.
From Eq.~(\ref{fd}) one can see that for $\theta<\pi\lambda$
the Born approximation renders an unphysical result, namely, $d_R>1$,
suggesting that this approximation is not applicable for small
$\theta$. This manifests the change in the mechanism of the
plasmon scattering which takes place for $\theta \lesssim
\lambda$. Indeed, at small $\theta$, incident wave travels closely
to the wave $d_R$ over a long distance, so that their amplitudes get
mutually redistributed. Importantly, in describing this
redistribution one can: ({\em i}) neglect both left-deflected,
$d_L$, and reflected, $r$, waves and ({\em ii}) employ
semiclassical approach, which yields
\begin{eqnarray}\label{incid}
&&t(k)=\cos\left(\frac {2\lambda}\theta\int_{0}^{\infty} dz
K_0\left[\sqrt{k^2d^2+ z^2}\right]\right),\\
&&d_R(k)=i\sin\left(\frac {2\lambda}\theta\int_{0}^{\infty} dz
K_0\left[\sqrt{k^2d^2+ z^2}\right]\right),\nonumber
\end{eqnarray}
where $K_0$ is the MacDonald function. A remarkable feature of
this result is that $t$ and $r$ {\it oscillate} strongly with
$\theta$, and that the oscillations {\it scale with the
interaction constant}. Note, that in the short-wavelength limit
$kd\gg1$, the perturbative result Eq.~(\ref{fd}) is valid even for
$\lambda>\theta$. Using the large-argument asymptote of $K_0(z)$,
it is easy to see that Eq.~(\ref{incid}) reproduces Eq.~(\ref{fd})
in this limit. For long-wavelengths, $kd\ll1$, Eq.~(\ref{incid})
yields $d_R=\sin(\pi \lambda/\theta)$, $t=\cos(\pi
\lambda/\theta)$, so that the perturbative and semi-classical
results match at $\lambda/\theta\lesssim 1$.

To outline the derivation of Eq.~(\ref{incid}), we note that the
system of equations, Eqs.~(\ref{ce1}), can be rewritten as two
independent closed equations,
\begin{eqnarray}
\label{pme}\left(\omega^2/v_{\scriptscriptstyle F}^2\right)u_{\pm}(x)=
\hat{D}\{u_{\pm}(x)\}\pm\hat{F}\{u_{\pm}(x)\},
\end{eqnarray}
where combinations $u_{\pm}(x)= u_1(x)\pm u_2(x)$ are introduced.
Searching for the semiclassical solution of Eq.~(\ref{pme}) in the
form $u_{\pm}(x)= e^{ikx+i\varphi_{\pm}(kx)}$, with slowly varying
phase, $\varphi_{\pm}^\prime\ll1$, we find
\begin{eqnarray}\label{viak}
\hspace{-0.3cm}2\varphi_\pm^\prime(kx)=\mp \lambda
K_0\left[\left(1+\varphi_\pm^\prime(kx)\right)
k\sqrt{d^2+x^2\theta^2}\right].
\end{eqnarray}
In evaluating the r.h.s. we assumed that $\theta$ is small. We see
that when $\lambda$ is small, the assumption,
$\varphi_{\pm}^\prime\ll1$, is justified. Then the smallness of
$\varphi_{\pm}^\prime$ allows one to neglect it in the argument of
$K_0$. Upon integrating Eq.~(\ref{viak}), we find $\varphi_{\pm}$.
Then transforming back to $u_1$ and $u_2$, we recover
Eq.~(\ref{incid}). The expression for $r(k)$ generalized to the
domain $\theta<\lambda<1$ follows from Eqs.~(\ref{fd}) and
(\ref{incid})
\begin{eqnarray}\label{qcrc}
r(k)\big |_{\theta<\lambda}&&\!\!\!\!\!\!=d_Rd_L\\
&&\!\!\!\!\!\!=\frac{\pi
\lambda}2e^{-\frac{2kd}\theta}\sin\left(\frac
{2\lambda}\theta\!\!\int_{0}^{\infty}\!\! dz
K_0\left[\sqrt{k^2d^2+ z^2}\right]\right),\nonumber
\end{eqnarray}
and in the long-wavelength limit simplifies to
$r|_{\theta<\lambda}=(\pi \lambda/2)\sin(\pi \lambda/\theta)$.

\noindent{\it Tunnel density of states}. Most importantly, the
non-trivial dependence of the plasmon scattering on
$\lambda$ and $\theta$ manifests itself in the {\it observables},
e.g., in the dependence of tunnel density of states, $\nu(\omega,
x)$, on the distance, $x$, from the intersection. To illustrate
this, consider first  a {\it single} nanotube with inhomogeneity
at $x=0$ which scatters plasmons with reflection coefficient
$\tilde{ r}(k)$. Then the correction to the tunnel density of
states reads
\begin{eqnarray}\label{ctdos}
&&\hspace{-1cm}\frac{\delta\nu(\omega,
x)}{\nu_0(\omega)}=\Gamma(\alpha+1)\sqrt{\alpha^2+\alpha/2}\\
&&\times\frac{|\tilde{r}(\omega/s)|}{\left(2\omega
x/s\right)^{\alpha+1}}\sin\left[\frac{2\omega
x}s-\varphi(\omega/s)-\frac{\pi\alpha}2\right],\nonumber
\end{eqnarray}
where $\varphi(k)=\arg(\tilde{r})$. Eq.~(\ref{ctdos}) follows from
the expression for interaction contribution to the local Green
function which takes into account the plasmon scattering,
\begin{eqnarray}\label{lGf}
&&\hspace{-0.4cm}\mathcal{G}(x,t)=\\
&&\hspace{-0.4cm}\exp\left\{\!-\pi\!\!\int
\!\!dk\left[\frac{|\int^x\!u_k|^2sk}{8v_{\scriptscriptstyle F}}
+\frac{|u_k(x)|^2v_{\scriptscriptstyle
F}}{8sk}\right]\!\!\left(1-e^{-is|k| t}\right)\!\right\}.\nonumber
\end{eqnarray}
Here $u_k(x)$ are the plasmon eigenmodes:  $u_k(x)=
(e^{ikx}+\tilde{r}(k)e^{-ikx})/\sqrt{2\pi}$, and
$u_k(x)=\sqrt{(1-|\tilde{r}|^2)/2\pi}\,e^{ikx}$, for $kx<0$ and
$kx>0$, respectively. Expanding the exponent in Eq.~(\ref{lGf})
with respect to $\tilde{r}$, and evaluating $\nu(\omega,
x)=\pi^{-1} \text{Re}\int_0^\infty dte^{i\omega
t}\mathcal{G}(x,t)$, we arrive at Eq.~(\ref{ctdos}).

Eq.~(\ref{lGf}) emerges upon representing electrons via the dual
bosonic fields $\theta_{i\alpha}$ and $\phi_{i\alpha}$,
$\psi_{i\alpha}\sim e^{i(\phi_{i\alpha}\pm\theta_{i\alpha})}$;
$i=1,2$ labels  the two bands, $\alpha=\uparrow, \downarrow$ are
the spins \cite{kane97}. Interaction is completely described by
the charged field
$\theta_c=\frac12\sum_{i\alpha}\theta_{i\alpha}$, while the three
neutral sectors are non-interacting. Expansion \cite{me}
$\theta_c(x)=\pi\sqrt{n_0}\int dk u_k(x)\hat{Q}_k $,
$\phi_c(x)=\frac1{\hbar\sqrt{n_0}} \int dk
[\int^x\!dyu_k(y)]\hat{P}_k $ over the plasmon eigenmodes,
$u_k(x)$, reduces the interacting Hamiltonian to a system of
harmonic oscillators $\{\hat{Q}_k,\hat{P}_k\}$ yielding
Eq.~(\ref{lGf}).

A simple reasoning allows to generalize Eq.~(\ref{ctdos}) to the
case of two intersecting nanotubes. Indeed, with intersection
playing the role of inhomogeneity, instead of one reflected wave
with reflection coefficient $\tilde{r}$ we have two independent
modes, $u_{\pm}(x)$, solutions of Eq.~(\ref{pme}), with reflection
coefficients $r_{\pm}$. It is important that while the absolute
values $r_+$ and $r_-$ are the same and equal to $|d_L|$, given by
Eq.~(\ref{fd}), their {\it phases } are different and are equal to
$\varphi_+(kx)- \varphi_+(-kx)$ and $\varphi_-(kx)-
\varphi_-(-kx)+\pi$, respectively, where $\varphi_\pm$ are
determined by Eq.~(\ref{viak}). Due to this difference in phases,
the oscillations $\propto\sin[2\omega x/s+\varphi]$ in
Eq.~(\ref{ctdos}) transform into a beating pattern
\begin{eqnarray}\label{mr}
&&\hspace{-0.4cm}\frac{\delta\nu(\omega,x)}{\nu_0(\omega)}=-
\Gamma(\alpha+1)\sqrt{\alpha^2+\alpha/2}\,\, \frac{\pi
\lambda}2\frac{
e^{-2\omega d/s\theta}}
{\left(2\omega x/s\right)^{\alpha+1}} \\
&&\hspace{-0.4cm}\times\sin\!\left(\!\frac
{2\lambda}\theta\!\!\int\limits_{0}^{\omega x\theta/s}\!\!\!\!\!
dz K_0\!\left[\sqrt{\left(\omega d/s\right)^2+
z^2}\right]\!\!\right)\!\cos\left(\frac{2\omega
x}s-\frac{\pi\alpha}2\right).\nonumber
\end{eqnarray}
Eq.~(\ref{mr}) is our main result. Remarkably, the {\it shape} of
the envelope of $\cos(2\omega x/s)$ oscillations depends strongly
on the interaction parameter, $\lambda$, offering a unique
signature of Luttinger liquid behavior. In particular, the number
of nodes in the envelope is equal to $\lambda/\theta$. Examples of
oscillations Eq.~(\ref{mr}) are plotted in Fig.~2 in terms of
tunnelling conductance, $G(V,x) \propto \nu(\omega=V,x)$, for
different interaction parameters. Note, that the language of
reflected plasmons, $r_+$, $r_-$, applies at distances $x\gg s/V$,
over which the reflection coefficient is formed.
Since the characteristic scale of the envelope is $s/V\theta$,
Eq.~(\ref{mr}) is valid as long as $\theta\ll1$. For large
$x\gg s/V\theta$, the argument of the sine in Eq.~(\ref{mr})
saturates at $\pi \lambda/\theta$. On the physical grounds,
the magnitude of the $\cos(2V x/s)$
oscillations at large $x$ should be given by Eq.~(\ref{ctdos}),
with element of scattering matrix $r$ instead of $\tilde{r}$. From
Eq.~(\ref{qcrc}) we realize that this is indeed the case.

\noindent{\it Implications}. Our main  prediction is that, for
purely {\em capacitive} coupling between nanotubes, the
conductance $G(V)$ into one or both ends of crossed nanotubes must
exhibit a structure, like shown in Fig.~2, with a large
characteristic "period" $V\sim s/(L\theta)$, where $L$ is the
distance from the end to the intersection. Smallness of $\theta$
insures that this structure (envelope in Fig.~2) is
distinguishable from size-quantization-like "filling" of the
envelope \cite{kane97,shahbazyan}, which changes with the period
$V=\pi s/L$. Also, an important prediction is that the envelope
beating structure in Fig.~2 vanishes with temperature much {\em
slower} than the filling, which vanishes at $T \sim s/L$.

The loop geometry of Ref. \onlinecite{refael} offers another
possible experimental implication. For this geometry, the easiest
way  to compare  the Sagnac oscillations in
Ref.~\onlinecite{refael} and our finding Eq.~(\ref{mr}) is to
assume that interaction is weak. Then the contribution to the
differential conductance from the Sagnac effect is roughly the
product of "size-quantization" oscillations, $\propto
\cos(2VL/v_F)$, and the envelope  $\propto \cos(2VLu_g/v_F^2)$,
where $V$ is the source-drain bias; $L$ is the loop perimeter and
$u_g$ is the gate-induced detuning of the "left" and "right"
velocities. Our Eq.~(\ref{mr}) for this geometry contains the same
first cosine $\cos(2VL/v_F)$, while the envelope is {\em entirely}
due to interactions. Thus, common feature of the two effects is
that envelopes survive at ``high'' temperatures when Fabry-Perot
oscillations vanish.

\noindent{\it Concluding remarks}. Adding a second parallel
nanotube to a given one leads \cite{matveev93} to a reduction of
$\alpha$ in $G(V)$ by  a factor of $2$. One could expect that for
a finite crossing angle, the effect of the second nanotube is
weaker. We found, however, that $G(V)$ depends on $\theta$ in a
{\em nonanalytical} fashion when $\theta \rightarrow 0$. This
nonanalyticity translates into a peculiar bias dependence of $G$,
as shown in Fig. 2. Thus, for crossed nanotubes, $G(V)$ is
extremely sensitive to the value of intratube Luttinger liquid
parameter, $g$. In armchair nanotubes, the currently accepted
value \cite{bockrath,yao99,postma} is in the range $0.19\div
0.26$. We emphasize that changing $g$ from $0.19$ to $0.26$ leads
to the increase of the interaction parameter, $\lambda$, by a
factor of $2$, which would have a drastic effect on the shape of
envelope in $\delta G(V)$, Fig.~2.

Concerning relation between our study and earlier studies
\cite{gao,crossedjunc} of crossed nanotube junctions, this
relation is exactly the relation between plasmon and electron
scattering. In the above papers anomalies were due to either
direct passage of electrons through the crossing point
\cite{crossedjunc} or due to crossing-induced {\em electron}
backscattering \cite{gao}. Scattering of plasmons was disregarded
in Ref.~\onlinecite{gao}. This is justified for perpendicular
nanotubes of Ref.~\onlinecite{gao}. As shown in our manuscript,
scattering of plasmons becomes important at small angles.

The work was supported by the  Petroleum Research Fund (grant
43966-AC10), DOE (grant DE-FG02-06ER46313) and by the Research
Corporation (JMG).

\end{document}